%% file: main.tex
\documentclass[sigconf,authorversion,nonacm]{acmart}

\usepackage{amsmath}

\usepackage{amsthm,amsfonts,amssymb}
\usepackage{multicol}
\usepackage{float,graphicx}
\usepackage{caption,subcaption}
\usepackage{url}
\usepackage[colorinlistoftodos]{todonotes}
\usepackage[skins,breakable]{tcolorbox}
\usepackage[utf8]{inputenc}
\usepackage{stmaryrd}
\usepackage{algorithm}
\usepackage{balance}

\input{macros}

\newcommand{\snote}[1]{{\color{brown}{\bf [SG: #1]}}}

\copyrightyear{} 
\acmYear{} 
\acmConference{}
\acmBooktitle{}
\acmPrice{}
\acmISBN{} 
\acmDOI{}

\title{Compressed Oblivious Encoding for Homomorphically Encrypted Search}
\titlenote{A preliminary version of this paper will appear at ACM CCS ’21. Authors are named alphabetically, and contributed equally.}

\author{Seung Geol Choi}
\affiliation{%
 \institution{United States Naval Academy}
 \country{}
}
\email{choi@usna.edu}

\author{Dana Dachman-Soled}
\affiliation{%
 \institution{University of Maryland}
 \country{}
}
\email{danadach@ece.umd.edu}

\author{S. Dov Gordon}
\affiliation{%
 \institution{George Mason University}
 \country{}
}
\email{gordon.dov@gmail.com}

\author{Linsheng Liu}
\affiliation{%
 \institution{George Washington University}
 \country{}
}
\email{lls@gwu.edu}

\author{Arkady Yerukhimovich}
\affiliation{%
 \institution{George Washington University}
 \country{}
}
\email{arkady@gwu.edu}

\setcopyright{none}
\begin{document}
\makeatletter
\def\@copyrightspace{\relax}
\makeatother
\fancyhead{}

\input{abstract}


\maketitle

\input{intro}

\input{prelim}

\input{coe}

\input{BFindex}

\input{powerSums}

\input{countingBF}

\input{ssp}

\input{exp}

\input{related}

\input{conclusion}
\input{Acknowledge}

\bibliographystyle{ACM-Reference-Format}
\balance
\bibliography{abbrev2,crypto,refs}

\input{app}
\end{document}

%% file: macros.tex
\newcommand{\e}{\epsilon}

\def\from{\leftarrow}

\def\Enc{{\sf Enc}}
\def\Gen{{\sf Gen}}
\def\Dec{{\sf Dec}}

\newcommand{\HHH}{\mathcal{H}}

\def\Encode{\mathsf{Encode}}
\def\Decode{\mathsf{Decode}}

\def\lbra{{\llbracket}}
\def\lket{{\rrbracket}}
\def\nil{{\mathsf{nil}}}

\renewcommand{\paragraph}[1]{\medskip \noindent{\bf #1}}

\newcommand{\negl}{\mathsf{negl}}
\newcommand{\zo}{\{0,1\}}
\newcommand{\idx}{\mathsf{nzx}}
\renewcommand{\to}{\rightarrow}

\newcommand{\Exp}{{\bf Exp}}

\newcommand{\mc}[1]{\mathcal{#1}}

\def\fhe{\mathsf{FHE}}
\def\ssp{\mathsf{sec\mbox{-}search}}
\def\Game{\mathsf{Game}_\fhe^\ssp(\mc{A})}
\def\Adv{\mathsf{Adv}_\fhe^\ssp(\mc{A})}

\newcommand{\A}{\mc{A}}

\newcommand{\BD}{\begin{definition}}
\newcommand{\ED}{\end{definition}}

\newcommand{\BCR}{\begin{corollary}}
\newcommand{\ECR}{\end{corollary}}

\newcommand{\BEX}{\begin{example}}
\newcommand{\EEX}{\end{example}}

\newcommand{\BL}{\begin{lemma}}
\newcommand{\EL}{\end{lemma}}

\newcommand{\BP}{\begin{proposition}}
\newcommand{\EP}{\end{proposition}}

\newcommand{\BCM}{\begin{claim}}
\newcommand{\ECM}{\end{claim}}

\newcommand{\BPF}{\begin{proof}}
\newcommand{\EPF}{\end{proof}}
\newcommand{\BEN}{\begin{enumerate}}
\newcommand{\EEN}{\end{enumerate}}

\newcommand{\BI}{\begin{itemize}}
\newcommand{\EI}{\end{itemize}}

\newcommand{\BO}{\begin{observation}}
\newcommand{\EO}{\end{observation}}

\newcommand{\BDS}{\begin{description}}
\newcommand{\EDS}{\end{description}}

\newcommand{\secparam}{\lambda}

%% file: abstract.tex
\begin{abstract}
Fully homomorphic encryption (FHE) enables a simple, attractive
framework for secure search.  Compared to other secure search systems,
no costly setup procedure is necessary; it is sufficient for the client
merely to upload the encrypted database to the server. Confidentiality
is provided because the server works only on the encrypted query and
records. While the search functionality is enabled by the full
homomorphism of the encryption scheme.

For this reason, researchers have been paying increasing attention to
this problem. Since Akavia et al. (CCS 2018) presented a framework for
secure search on FHE encrypted data and gave a working implementation
called SPiRiT, several more efficient realizations have been proposed. 

In this paper, we identify the main bottlenecks of this framework and
show how to significantly improve the performance of FHE-base secure
search. In particular,  
\BI
\item To retrieve $\ell$ matching items, the existing framework needs to
repeat the protocol $\ell$ times sequentially. In our new framework, all
matching items are retrieved in parallel in a {\em single protocol
execution}.

\item The most recent work by Wren et al. (CCS 2020) requires $O(n)$
multiplications to compute the first matching index. Our solution
requires {\em no homomorphic multiplication}, instead using only
additions and scalar multiplications to encode all matching indices.  

\item 
  Our implementation and experiments show that to fetch 16 matching
  records, our system gives an 1800X speed-up over the state of the art
  in fetching the query results resulting in a 26X speed-up for the full
  search functionality.

\EI

%
\end{abstract}

%% file: intro.tex
\section{Introduction}

As computing paradigms are shifting to cloud-centric technologies, users
of these technologies are increasingly concerned with the privacy and
confidentiality of the data they upload to the cloud. Specifically, a
\emph{client} uploads data to the \emph{server} and expects the
following guarantees: 

\BEN
\item The uploaded data should remain private, even
from the server itself;
\item The server should be able to perform computations on the uploaded
data in response to client queries;
\item The client should be able to efficiently recover the results of
the server's computation with minimal post-processing.
\EEN

In this work, we will focus on the computational task of secure search.
In this application, the client uploads a set of records to the server,
and later posts queries to the server.  Computation proceeds in two
steps called \emph{matching} and \emph{fetching}.
In the matching step, the server compares the encrypted search query from
the client with all encrypted records in the database, and computes an
encrypted 0/1 vector, with 1 indicating that the corresponding record
satisfies the query.  The fetching step returns all the 1-valued
indexes and the corresponding records, to the client for decryption.

While seemingly conflicting goals, the guarantees of (1), (2), (3) can
be simultaneously achieved for the secure search setting via techniques
such as secure multiparty computation and searchable encryption.
Recently, a line of works has focused on Fully Homomorphic Encryption
(FHE)-based secure search, which we describe next.

\paragraph{FHE-based secure search.}
The simplicity of the framework of {\em secure search on FHE encrypted
data} is attractive. Compared to other secure search systems, no costly
setup procedure is necessary; it is sufficient for the client merely to
upload the encrypted database to the server. Confidentiality is provided
because the server works only on the encrypted query and records. The
server can still perform the search correctly due to the powerful
property of the full homomorphism of the underlying encryption scheme. 

\begin{figure}
\includegraphics[width=3.5in]{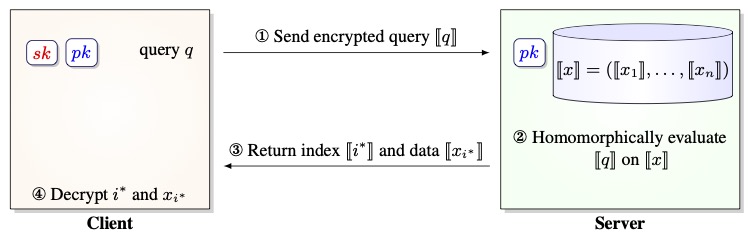}

\hspace{-4em}{\footnotesize In the above, $\lbra \cdot \lket$ denotes
an FHE-encrypted ciphertext.}

\caption{The secure search framework in~\cite{CCS:AkaFelSha18}}
\label{fig:secure-search}
\end{figure}

For this reason, researchers have been paying increasing attention to
this problem. In particular, Akavia et~al.~\cite{CCS:AkaFelSha18}
introduce a framework of performing secure search on FHE-encrypted data
(see Figure~\ref{fig:secure-search}). 

Informally, a secure, homomorphic encrypted search scheme has the
following Setup: 
\BEN
\item (Setup) The client encrypts and uploads $n$ items $x = (x_1, \ldots, x_n)$
to the server. Let $\lbra x \lket = (\lbra x_1
\lket, \ldots, \lbra x_n \lket).$ denote the encrypted data stored in
the server. 
\EEN

Throughout the paper, we let $\lbra \cdot \lket$ denote an FHE-encrypted
ciphertext. After the encrypted records have been uploaded, the client
can perform a secure search using three algorithms, (Query, Match,
Fetch).
\BEN
\setcounter{enumi}{1}
\item (Query) The client sends an encrypted query $\lbra q \lket$ to the server.  

\item (Match) The server homomorphically evaluates the query $\lbra q \lket$ on
each record $\lbra x_i \lket$ to obtain the encrypted matching
results $\lbra b \lket = (\lbra b_1 \lket, \ldots, \lbra b_n \lket).$
That is, $b_i$ is 1 if item $x_i$ satisfies the given query $q$;
otherwise, $b_i$ is $0$.

\item (Fetch) Given $\lbra b \lket$, the server homomorphically computes
$\lbra i^* \lket$, where $i^* = \min \{ i \in [n] : b_i = 1 \}$ which
corresponds to the first matching record index. It fetches  $\lbra
x_{i^*} \lket$ (obliviously) and sends $(\lbra i^* \lket, \lbra x_{i^*}
\lket)$ to the client for decryption. 
\EEN

\paragraph{Multiplications in the fetching step.}
Akavia et al.~also provide a construction that performs the fetching
step in $O(n \log^2 n)$ homomorphic multiplications. Subsequently, more
efficient algorithms have been presented with $O(n \log n)$
multiplications~\cite{PETS:AGHL19} and $O(n)$
multiplications~\cite{CCS:WYXZ20}.

\subsection{Motivation}
\paragraph{Bottleneck: fetching records sequentially.}
Suppose a client wants to fetch all matching items.  Under the above
framework, the client would first obtain the first matching index $i^*$
and its corresponding item $x_{i^*}$.  To fetch the second matching
item, the framework suggests that the client should slightly change the
original query $q$ to a new query $q'_{i^*}$ as follows:

\BI
\item $q'_{i^*}(i, x_i)$ return true if $q(i, x_i)$ is true and $i > i^*$. 
\EI

Then, by executing a new instance of the protocol with the encrypted
query $\lbra q'_{i^*} \lket$, the client will obtain the second matching
item.  By repeating this procedure, the client will ultimately obtain
all the matching records.  

Note that the query $q'_{i^*}$ embeds $i^*$ in itself as a constant,
which implies that there is no way for the client to construct this
query $q'_{i^*}$ without obtaining $i^*$ first. In other words, the
client can construct the query for the second matching item, {\em only
after fetching the first matching item}. In this sense, the framework
inherently limits the client to fetch only a single matching record at a
time in a sequential manner. 

If there are $\ell$ matching records, the client and server have to
execute $\ell$ instances of the Query, Match, and Fetch algorithms.
Since each Match and Search step requires costly homomorphic
multiplications, the limitation of sequential protocol execution creates
a serious bottleneck with respect to the running time. This leads us to
ask the following natural question:

\medskip

\BI
\item[]
\it
Is there a different secure search framework that allows the client to
fetch all the matching records by executing a smaller number of protocol
executions, possibly avoiding sequential record fetching?
\EI

\paragraph{Reducing homomorphic multiplications.} 
All previous schemes have to perform $\Omega(n)$ homomorphic
multiplications in the fetching step. Since homomorphic multiplications
are costly operations, it is desirable to reduce such computations,
which begs the natural following question: 

\medskip
\BI
\item[]
\it
Can you reduce the number of homomorphic multiplications in the fetching
step? 
\EI
In this paper, we answer both of the above questions affirmatively. 

\begin{figure*}
\begin{tabular}{lccccccc}
\hline 
  & rounds & \#Match & ${\sf hmult}$ & ${\sf hadd}$ & ${\sf smult}$ & communication  & plaintext modulus \\
\hline 
LEAF~\cite{CCS:WYXZ20}    & $s$ & $s$ & $O(ns)$   &  $O(ns\log n)$     & 0      & $O(s \cdot \log n \cdot |C|)$ & 2 \\ 
Protocol w/ BF-COIE   & $3$ & 1 &  0 & $O(n \log \frac n s)$ & 0               & $O(s^{1+\e} \log \frac n s \cdot |C| + pir(s))$ & prime \\
Protocol w/ PS-COIE   & $3$ & 1 & 0 & $n \cdot s$   &  $n \cdot s$                  & $O(s \cdot |C| + pir(s) )$ & prime \\ 
Protocol w/ BFS-CODE  & $2$ & 1 & $n$    &    $O(\secparam n)$ &   0        & $O(s \secparam \cdot |C|)$ & prime \\ 
\hline
\end{tabular}

\BI
\item $\secparam$: statistical security parameter.
\item $n$: number of uploaded encrypted records.
\item $s$: number of matching records. 
\item $\e$: protocol parameter such that $0 < \e < 1$.
\item \#Match: number of times the matching algorithm is executed. 
\item ${\sf hmult}$: number of homomorphic multiplication operations
used in the overall fetching step.
\item ${\sf hadd}$: number of homomorphic addition operations used in the overall fetching step. 
\item ${\sf smult}$: number of scalar (plain) multiplication
operations used in the overall fetching step.
\item $|C|$: length of an FHE ciphertext.
\item $pir(s)$: communication complexity required to retrieve $s$ records via a PIR protocol. 
\EI

\caption{Performance Comparisons when $s$ records are fetched\label{fig:comparison}}
\end{figure*}

\subsection{Our Work}
\paragraph{Parallelizing the Fetch procedure.}
To address the issues, we introduce a new secure search framework where
the matching items are retrieved {\em in parallel} in a constant number
of rounds.  Our Setup, Query and Match algorithms are the same as in
prior work.  However, we modify the Fetch 
procedure, dividing into two steps:~Encode
and Decode.  In the Encode step, the server 
homomorphically inserts the matching items 
into a data structure - the particular structure depends on the construction, as
we provide 3 different constructions, each using a different encoding.  
After receiving the encrypted encoding, the 
client decrypts the encoding and runs the Decode step to recover the items.  

\paragraph{Compressed oblivious encoding.}
The encoding is computed homomorphically, and, most importantly, allows
to encode {\em the full result set}, rather than just a single item.  
In particular, we introduce a notion of Compressed Oblivious Encoding (COE).  A compressed oblivious encoding takes as input a large, but sparse, vector and compresses it to a much smaller encoding from which the non-zero entries of the original vector can be recovered.  What makes this encoding \emph{oblivious} is that the encoding procedure is performed on encrypted data.  
In certain constructions, the encoding includes the data values (CODE, compressed oblivious data encoding), and in
others it only includes the indices (COIE, compressed oblivious index encoding).  In the latter case, the Decode
procedure is interactive, and allows the client to recover the values
from the decoded set of indices.  

For simplicity, when describing the 
generic syntax of secure search scheme, we denote the Encode procedure as taking both the
indices and the values as input, and we suppress the fact that when the
values are not used during Encoding, the Decoding step must be
interactive. Recall, we use $\lbra b \lket = (\lbra b_1\lket, \ldots, \lbra b_n \lket)$
to denote the encrypted bit vector that results from the Match step. 

\BEN
\setcounter{enumi}{3}
\item (Encode) Let $S = \{ i \in [n] : b_i = 1 \}$.  Let $V = \{v_i : i \in S\}$. The server
homomorphically evaluates an $\lbra \mathsf{encoding}(S,V) \lket$ and send it to
the client. 
\item (Decode) The client decrypts $\lbra
\mathsf{encoding}(S) \lket$ and runs the decoding 
procedure to recover $(S, V)$.
\EEN

We assume that the results set $|S|$ is small (i.e., sublinear in $n$).
We would like the size of the compressed encoding to be sublinear in $n$ to
maintain meaningful communication cost. 

\paragraph{No multiplications in the Encode step.} 
To ensure minimal computational cost for encoding the results, we also wish to
minimize the number of homomorphic multiplications.  Recall, the best prior
work requires $O(n)$ multiplications by the server.  Somewhat surprisingly, we
demonstrate three encoding algorithms that can be evaluated {\em without} any
homomorphic multiplications!

\paragraph{Using PIR (Private Information Retrieval).} 
The asymptotic complexities and trade-offs of the search protocols are presented in
Figure~\ref{fig:comparison}.

In some of our
protocols (i.e., the search protocols with BF-COIE and PS-COIE; see Sections 4 and 6.3 for more detail), the indices and actual records are fetched in separate steps.  This allows us to focus on optimizing the retrieval of the indices after which the values can be fetched using an efficient (setup-free) PIR protocol resulting in overall savings.  

However, if reliance on PIR is undesirable, we also offer a variant that fetches the values directly (i.e., the protocol w/ BFS-CODE in Figure~\ref{fig:comparison}; see Sections 5 and 6.4 for more detail), as in prior work.


\paragraph{Implementation.} We implement all of our proposed schemes and
compare their performance with that of prior work. Our experiments show
that our schemes outperform the fetching procedure of prior work by a
factor of 1800X when fetching 16 records, which results in a 26X speedup
for the full search functionality.

%% file: prelim.tex
\section{Preliminaries}

Let $\secparam$ be the security parameter. 
%
%
For a vector $a$, let $\idx(a)$ denote the set of all the positions $i$
such that $a_i$ is non-zero, i.e., $$\idx(a) := \{ i: a_i \ne 0\}.$$  

\paragraph{Chernoff bound.} We will use the following version of
Chernoff bound. 

\begin{theorem}
Let $X_1, \ldots, X_n$ be independent random variables taking values in
$\zo$ such that $\Pr[X_i = 1] = p$. Let $\mu := \Exp[\sum X_i] = np$.
Then for any $\delta > 0$, it holds
$$\Pr\left[ \sum_{i=1}^n X_i \ge (1 + \delta) \mu \right] \le 
\left (\frac {e^\delta} {(1+\delta)^{(1+\delta)}} \right)^\mu.$$ 
\end{theorem}

\paragraph{FHE.} We use a standard CPA-secure (leveled) fully
homomorphic encryption scheme $(\Gen, \Enc, \Dec)$.  We refer readers
to~\cite{PETS:AGHL19,CCS:WYXZ20} for a formal definition.  We use $\lbra
x \lket$ to denote an encryption of $x$. 

We also use $+$ (resp. $\cdot$) to denote homomorphic addition (resp.,
multiplication). For example, $\lbra c \lket := \lbra a \lket + \lbra b
\lket$ means that homomorphic addition of two FHE-ciphertexts $\lbra a
\lket$ and $\lbra b \lket$ has been applied, which results in $\lbra c
\lket$. 

\paragraph{PIR.} A PIR protocol allows the client to choose the index
$i$ and retrieve the $i$th record from one (or more) untrusted server(s)
while hiding the index value $i$~\cite{JACM:CKGS98}.  

Assume that each of the $k$ server has $n$ records $D = (d_1, \ldots, d_n)$
where all items $d_i$ have equal length. A single-round $k$-server PIR
protocol consists of the following algorithms:
\BI
\item The query algorithms $Q_j(i, r) \to q_j$ for each server $j \in [k]$, which
are executed by the client with input index $i$ and randomness $r$. 
\item The answer algorithms $A_j(D, q_j) \to a_j$ for each server $j \in
[k]$, which is executed by the $j$th server. 
\item The reconstruction algorithm $R(i, r, (a_1, \ldots, a_k)) \to d_i$. 
\EI
The communication complexity of a PIR protocol is defined by the sum of
the all query lengths and answer lengths, i.e.,  
$$\sum_{j \in [k]} |q_j| + |a_j|.$$

A PIR protocol is correct if for any $D = (d_1, \ldots, d_n)$ with $|d_1| =
\cdots = |d_n|$, and for any $i \in [n]$, it holds that 
    $$\Pr_r \bigg [ R \Big(i, r, \big \{ A_j(D, Q_j(i,r)) \big \}_{j=1}^k \Big ) = d_i \bigg ] = 1.$$
A PIR protocol is private if for any $j \in [k]$, for any $i_0, i_1 \in
[n]$ with $i_0 \ne i_1$, the following distributions are computationally
(or statistically) indistinguishable: 
  $$\{Q_j(i_0, r)\}_r \approx \{Q_j(i_1, r)\}_r.$$

\subsection{Bloom Filter}
A Bloom filter~\cite{bloom70spacetime} is a well-known space-efficient
data structure that allows a user to insert arbitrary keywords and later
to check whether a certain keyword in the filter. 

\paragraph{$\sf BF.Init()$.}
The filter $B$ is essentially an $\ell$-bit vector, where $\ell$ is a
parameter, which is initialized with all zeros.
The filter is also associated with a set of $\eta$ different hash
functions $$\HHH = \{ h_q: \zo^* \to [\ell] \}_{q=1}^\eta.$$

\paragraph{$\sf BF.Insert(B, \alpha)$.}
To insert a keyword $\alpha$, the hash results are added to the
filter. In particular, 
  \BI
  \item For $q \in [\eta]$ do the following:
    \BI
    \item[] 
    Compute $j = h_q(\alpha)$ and set $B_j := 1$. Here $B_j$ is the $j$th bit of $B$.  
    \EI
  \EI

\paragraph{$\sf BF.Check(B, \beta)$.}
To check whether a keyword $\beta$ has been inserted to a BF filter
$B$, one can just check the filter with all hash results. In particular, 
  \BI
  \item For $q \in [\eta]$ do the following: 
    \BI
    \item[] Compute  $j = h_q(\beta)$ and check if $B_j$ is set. 
    \EI
  \item If all checks pass output "yes". Otherwise, output "no".
  \EI

The main advantage of the filter is that it guarantees there will be no
false negatives and allows a tunable rate of false positives: 
$$
\bigg(1 - \Big (1 - \frac{1}{ \ell} \Big )^{\eta s} \bigg)^\eta
\approx 
\Big(1 - e^{-\frac{\eta s}{ \ell}} \Big)^\eta,
$$
where $s$ is the number of keywords in a Bloom filter. 

\paragraph{Random oracle model for hash functions.}
We show our analysis in the random oracle model. That is, the hash
functions are modelled as random functions. 

\subsection{Algebraic Bloom Filter}
\label{sec:ABF}
In this work, we leverage a variant of the Bloom filter where, when
inserting an item, the bit-wise OR operation is replaced by addition.
There have been works using a similar idea of having each cell hold an
integer instead of holding a bit~\cite{TON:FCAB00, PODC:Mitzenmacher01}. 

Moreover, we consider a limited scenario {\em where the upperbound on the
number of keywords to be inserted is known beforehand}. In particular,
let $s$ denote such an upperbound. 

As before, the filter is also associated with a set of $\eta$ different
hash functions  $\HHH = \{ h_q: \zo^* \to [\ell] \}_{q=1}^\eta$.
However, now the filter $B$ is not an $\ell$-bit vector but a vector
where each element is in $[s \eta]$ (i.e., $B \in
[s\eta]^\ell$)~\footnote{We can reduce $s \eta$ further to $\Theta( \eta \cdot
(s/\ell) \cdot \log (s/\ell))$ using a Chernoff bound to bound the number of collisions contributing to the sum, but we will use $s\eta$ for
the sake of simplicity of presentation.}.
Therefore, the number of bits to encode $B$ is now blown up by
a multiplicative factor $\lceil \lg s\eta \rceil$. 

The BF operations are described below where differences are marked by
framed boxes. 

\paragraph{$\sf BF.Insert(B, \alpha)$.}
To insert a keyword $\alpha$, the hash results are added to the
filter. In particular, 
  \BI
  \item For $q \in [\eta]$ do the following:
    \BI
    \item[] 
    Compute $j = h_q(\alpha)$ and set \fbox{$B_j := B_j + 1$}. 
    \EI
  \EI

\paragraph{$\sf BF.Check(B, \beta)$.}
To check whether a keyword $\beta$ has been inserted to a BF filter
$B$, one can just check the filter with all hash results. In particular, 
  \BI
  \item For $q \in [\eta]$ do the following: 
    \BI
    \item[] Compute  $j = h_q(\beta)$ and check if $B_j$ is \fbox{greater than 0}. 
    \EI
  \item If all checks pass output "yes". Otherwise, output "no".
  \EI

It is easy to see that this variant construction enjoys the same
properties as the original BF construction.

%% file: coe.tex
\section{Compressed Oblivious Encoding}
As our main building block, we introduce a new tool we call Compressed Oblivious Encoding.  A compressed oblivious encoding takes as input a large, but sparse, vector and compresses it to a much smaller encoding from which the non-zero entries of the original vector can be recovered.  What makes this encoding \emph{oblivious} is that the encoding procedure is oblivious to the original data; in fact, in our constructions the original data will all be encrypted.  An efficient encoding must satisfy the following two performance requirements: 1) The size of the encoding must be sublinear in the size of the original array, and 2) constructing the encoding should be computationally cheap. Our constructions only use (homomorphic) addition and multiplication by constant (i.e. plaintext values).

A related notion is that of compaction over encrypted data~\cite{eprint:BlaAgu11,EC:AKLNPS20} which aims to put all non-zero entries of a vector to the front of the encoding.  Our encoding can be viewed as a form of noisy compaction where, in addition to keeping all the non-zero entries, it allows a small number zero entries to be mixed in with the result.  Thus, a compressed encoding trades some inaccuracy in the output for much cheaper construction costs.

We define two variants of compressed oblivious encodings, one that encodes the indices of non-zero entries and one that encodes the actual entries themselves.

\subsection{Compressed Oblivious Index Encoding}
A compressed oblivious index encoding (COIE) encodes the indices or locations of all the non-zero entries in the input array.  We begin by defining the parameters and syntax for a COIE scheme.

\paragraph{Parameters.} A COIE scheme is parametrized as follows. 

  \BI 
  \item $n$: Input size -- The dimension of the input vector $v$.
  \item $s$: Sparsity -- Bound on the number on non-zero entries in $v$.
  \item $c$: Compactness -- The dimension of the output encoding.
  \item $f_p$: False positives -- The upperbound on the number of false
  positives returned by the decoding algorithm.
  \EI

\paragraph{Syntax.} A $(n, s, c, f_p)$-COIE scheme has the following syntax: 
\BI
\item $\lbra \gamma_1 \lket, \ldots, \lbra \gamma_c \lket \from
\Encode(\lbra v_1 \lket, \ldots, \lbra v_n \lket)$.
The $\Encode$ algorithm takes as input a vector of ciphertexts with $v_i \in \zo$ for all $i \in [n]$.
It outputs an encrypted encoding $\lbra \gamma_1 \lket, \ldots, \lbra \gamma_c \lket$.

\item $I \from \Decode(\gamma_1, \ldots, \gamma_c)$. 
The $\Decode$ algorithm takes the encoding $(\gamma_1, \ldots,
\gamma_c)$, in decrypted form, and outputs a set $I \subseteq [n]$ 
\EI

\paragraph{Correctness.} 
Let $(\gamma_1, \ldots, \gamma_c) \from \Dec(\lbra \gamma_1 \lket,
\ldots, \lbra \gamma_c \lket)$ denote a correct decryption of
the encoding. 

\BD
A $(n, s, c, f_p)$-COIE scheme is \emph{correct}, if the following conditions
are satisfied: 
\BI
\item (No false negatives) For all $v \in \zo^n$ with at most $s$
non-zero positions, and for all $i \in \idx(v)$, it should hold 
 $$i \in \Decode( \Dec(\Encode(\lbra v_1 \lket, \ldots, \lbra v_n \lket)))$$ with probability at least $1-\negl(\secparam)$
where the random coins are taken from $\Encode$.

\item (Few false positives) For all $v \in D^n$ with at most $s$ non-zero positions, consider
the set of false positives $$E = \{i \in [n]: v_i = 0\mbox{, but } i \in I
\},$$
where $I = \Decode(\Dec(\Encode(\lbra v_1 \lket, \ldots, \lbra v_n \lket))).$

We require that $|E| \le f_p$ with the overwhelming probability over the
randomness of $\Encode$.  
\EI
\ED

\paragraph{Efficiency.} 
For efficiency, we look at the following three parameters of a COIE: 

\BI
\item The type and number of operations used by the $\Encode$ algorithm. 

\item The size of the encoding.

\item The computation cost of the $\Decode$ algorithm.
\EI

For an efficient construction, we require that the latter two of these are sublinear in the size of the input vector.






\subsection{Compressed Oblivious Data Encoding}
A Compressed Oblivious Data Encoding (CODE) scheme is very similar to
COIE except, rather than encoding the locations of non-zero entries, it encodes the values of these entries. We give a
definition of CODE below where differences are marked by framed boxes.

\paragraph{Parameters.} A CODE scheme is parametrized by the same four
parameters $(n,s,c,f_p)$ as a COIE.

\paragraph{Syntax.} A $(n, s, c, f_p)$-CODE scheme over \fbox{domain $D$} has the following syntax: 
\BI
\item $\lbra \gamma_1 \lket, \ldots, \lbra \gamma_c \lket \from
\Encode(\lbra v_1 \lket, \ldots, \lbra v_n \lket)$.
The $\Encode$ algorithm takes as input a vector of ciphertexts with $v_i \in \boxed{D}$ for all $i \in [n]$.
It outputs an encrypted encoding $\lbra \gamma_1 \lket, \ldots, \lbra \gamma_c \lket$.

\item $\boxed{V} \from \Decode(\gamma_1, \ldots, \gamma_c)$. 
The $\Decode$ algorithm takes the encoding $(\gamma_1, \ldots,
\gamma_c)$, in decrypted form, and outputs a set of values 
\fbox{$V = \{v_i : v_i \ne 0\}$}
\EI

\paragraph{Correctness.} 
\BD
A $(n, s, c, f_p)$-CODE scheme over domain $D$ is correct, if the following conditions
are satisfied: 
\BI
\item (No false negatives) For all $v \in \zo^n$ with at most $s$
non-zero positions, and for all $i \in \idx(v)$, it should hold 
 $$  \boxed{v_i \in \Decode( \Dec(\Encode(\lbra v_1 \lket, \ldots, \lbra v_n \lket)))} $$ 
 
 with probability $1-\negl(\secparam)$ where the random coins are taken from $\Encode$. 

\item (Few false positives) For all $v \in D^n$ with at most $s$ non-zero positions, consider
the set of false-positive values $$\boxed{E = \{ z \in V: z \ne v_i \mbox{~for any~} i \in \idx(v)\}},$$
where $V = \Decode(\Dec(\Encode(\lbra v_1 \lket, \ldots, \lbra v_n \lket))).$

We require $|E| \le f_p$ with the overwhelming probability over the randomness of $\Encode$.
\EI
\ED

%% file: BFindex.tex
\section{COIE Schemes }
We assume the input index vector $v \in \zo^n$ is sparse. In particular,
throughout the paper, we assume $s = o(n)$.

\subsection{A Warm-up construction}
Using an algebraic BF, we can create an $(n, s, c, f_p)$-COIE scheme
(the parameters $c$ and $f_p$ will be worked out after the description of
the scheme).

\paragraph{$\Encode(\lbra v_1 \lket, \ldots, \lbra v_n \lket)$.}
The encoding algorithm works as follows:

\BEN
\item Initialize a BF $\lbra B \lket := (\lbra B_1 \lket, \ldots, \lbra
B_c \lket)$ with $B_j = 0$ for all $j$. Let $\HHH = \{ h_q: \zo^* \to
[c] \}_{q=1}^\eta$ be the associated hash functions. 

\item For $i = 1, \ldots, n$:
    \BEN
    \item For $q = 1, \ldots, \eta$, do the following: 
      Compute $j = h_q(i)$ and set $\lbra B_j \lket := \lbra B_j \lket +
      \lbra v_i \lket$.
 
\EEN
\EEN

Note that at step 2.a in the above, if $v_i = 0$, then $B_j$ stays the
same. On the other hand, if $v_i = 1$, then $B_j$ will be increased by
1.  This implies that $B$ will exactly store the results of the
operations $\{ {\sf BF.Insert}(B, i) : i \in \idx(v)\}.$

\paragraph{$\Decode(B_1, \ldots, B_c)$.}
Given the algebraic BF $B$, we can recover
the indices for the nonzero elements as follows:
\BI
\item Initialize $I$ to be the empty set. 
\item For $i \in [n]$: if ${\sf BF.Check}(B, i)$ = ``yes", add $i$ to $I$.  
\item return $I$. 
\EI

\paragraph{Parameters $c$ and $f_p$.} Since this is a warm-up
construction, we perform only a rough estimation on the false positive parameter
and the compactness parameter. 

For reasons that will become clear later, we wish to keep the upper bound
on the number of false positives ($f_p$) small.  In
particular, we use a BF with false-positive rate $1/n$. Since there
are $n$ operations of $\sf BF.Check$, the expected number of false
positives is 1, and from the Chernoff bound, the number of false
positives is bounded by $\Omega(\log \secparam)$ with overwhelming
probability in $\secparam$. This implies that we have $f_p = \Omega(\log
\secparam).$

The dimension $c$ of the Bloom filter $B$ can be
computed using the following equation of BF false positive ratio: 

$$ \Big(1 - e^{-\frac{\eta s}{c}} \Big)^\eta \le \frac 1 n,$$ 

Setting $c = \eta s \cdot n^{\frac 1 \eta}$ will satisfy the
equation. This can be verified by using an equality $1 - e^{-x} \le x$
for $x \in [0,1]$; that is, $1 - e^{-\frac{\eta s}{c}} \le \frac{\eta
s}{c} = 1/n^{1/\eta}.$

\paragraph{Efficiency.}
\BI
\item The encoding algorithm uses $n\eta$ homomorphic addition
operations, and $n\eta$ hash functions.  

\item The dimension $c$ of the encoding is $\eta s \cdot n^{\frac 1
\eta}$. Usually, $\eta$ is set to between 2 and 32.

\item The decoding algorithm uses $n$ operations of $\sf BF.Check$.
\EI

In summary, we have reduced the encoding size $c$ to be sub-linear in $n$ as desired. However, we still need to reduce the number $\sf
BF.Check$ operations in Decode to be sub-linear in $n$. We show how to achieve that in our next construction.

\subsection{BF-COIE}\label{sec:BF-COIE}
We now show how to improve the above construction to achieve decoding in
time $o(n)$.  The main idea of this improvement is to use Bloom filters
to represent a binary search tree, one BF per level of the tree.  We can
then guide the decoding algorithm to avoid decoding branches that do not contain
non-zero entries.  As most branches can be truncated well before
reaching the leaf-level Bloom filter, this results in sublinear total
cost.


\paragraph{Example.} 
Before presenting the formal protocol for this construction we convey our idea through an example.  Let $n =
32$, and suppose we wish to encode the indices $I =  \{1, 15, 16\}$. 
Denote $$I^k = \left \{ \Big \lceil \frac i {2^k} \Big \rceil : i \in
I \right \}.$$

Intuitively, an element $i$ in $I^k$ can be thought of a range of length
$2^k$ covering $[(i-1)\cdot 2^{k}+1, i\cdot2^k]$. We have:
\BI
\item $I^4 = \{1\}.$
\item $I^3 = \{1, 2\}.$
\item $I^2 = \{1, 4\}.$
\item $I^1 = \{1, 8\}.$
\item $I^0 = \{1, 15, 16\}.$
\EI

Now, assume we insert each set $I^k$ into its own BF. We can traverse these BF's to decode the set $I$ as follows:

\BEN
\item Check $I^4$ for all possible indices. The only possible indices at this level are $1$ and $2$, since $n=32$ and $I^4$ divides the original indices by
$2^4 = 16$. 

In the above example, When we query the BF for $I^4$, it 
only contains the index $1$, which means that no values greater than 16 are contained in $I$.  We can thus avoid checking any such indices at the lower levels.  

Now consider the BF at the next level (i.e., the BF for $I^3$).  The
only possible values at this level are 1,2,3,4, but since we already
know that there are no values greater than 16 in $I$, we only need to
check for values $1, 2$ (since $3 \cdot 8 > 16$).

\item Check $I^3$ for indices $1, 2$. The BF will show that indices $1$ and $2$
are both present, which means that we need to check indices $1,2$ and $3,4$ in $I^2$. 
\item Check $I^2$ for indices $1, 2, 3, 4$. The BF will show that indices $1$ and $4$
are present, which means that we only need to check indices $1, 2$ and $7, 8$ in $I^1$, all other indices can be skipped.

\item Check $I^1$ for indices $1, 2, 7, 8$. The BF will show that indices $1$ and $8$
are present, which means that we need to check indices $1, 2$ and $15, 16$.
\item Check $I^0$ for indices $1, 2, 15, 16$, and output the
final present indices $1, 15, 16$. 
\EEN 

Assuming, for now, that there are no false positives, observe that this approach checks at most $2 \cdot |I|$ values at each
level, and there are $\lg n$ levels. Therefore, the decoding algorithm
will check $O( |I| \cdot \lg n)$ indices, which is sub-linear in $n$.  

\paragraph{BF-COIE.} We now describe our BF-COIE construction. As
before, we will work out the parameters after describing our
construction. The encoding algorithm is described in Algorithm~\ref{alg:BFCOIEE}.   

\begin{algorithm}[htbp]

\BEN
\item[] \hspace{-2em} For simplicity, $n$ and $s$ are assumed to be powers of 2. 

\item $t := \lg {\frac n {2s}}$ 
\item For $k = 0, \ldots, t$: 

\BEN
\item Initialize $\lbra B^k \lket = (\lbra B^k_1 \lket, \ldots, \lbra
B^k_\ell \lket) := (\nil, \ldots, \nil)$. 

\item Choose $\HHH^k = \{ h^k_q: \zo^* \to [\ell] \}_{q=1}^\eta$ at random.

\item For $i \in [n]$ and for $q \in [\eta]$: \\ 
  \BEN
  \item[] $i' := \lceil i/2^k \rceil$, 
        $j := h^k_q(i')$, \\
        If $\lbra B^k_j \lket$ is $\nil$, then $\lbra B^k_j \lket := \lbra v_{i'} \lket$ \\
        Otherwise, $\lbra B^k_j \lket := \lbra B^k_j \lket + \lbra v_{i'} \lket$
  \EEN
\EEN

\item Output $\lbra B^0 \lket, \ldots, \lbra B^t \lket$.
\EEN

\caption{ ${\sf BF\mbox{-}COIE}.\Encode(\lbra v_1 \lket, \ldots, \lbra v_n \lket)$}
\label{alg:BFCOIEE}
\end{algorithm}

Note that in steps (a) to (c) above, the warm-up construction is used to
construct BF $B^k$ for indices $I^k$. 

In order to reduce the size of the output encoding, we set $t$ to be
$\lg {\frac n {2s}}$ instead of $\lg n$ as described previously. Note
that when $t$ is set in this way, $I^t$ contains at most $n/2^t = 2s$
possible values thus maintaining our invariant.


The decoding algorithm is described in Algorithm~\ref{alg:BFCOIED}.   

\begin{algorithm}[htbp]
\BEN
\item Initialize $I, I^0, \ldots, I^{t-1} := \emptyset$
\item Initialize $I^t := \{1, \ldots, n/2^t\} = [2s]$
\item For $k = t, t-1, \ldots, 1$, and for $i' \in I^k$: \\
  $~~~$ If ${\sf BF.Check}(B^k, i')$ is ``yes", add $2i'-1$, $2i'$ in $I^{k-1}$  
\item For $i \in I^0$: \\
  $~~~$ If ${\sf BF.Check}(B^0, i)$ is ``yes", add $i$ to $I$  
\item Output $I$ 
\EEN
\caption{ ${\sf BF\mbox{-}COIE}.\Decode(B^0, \ldots, B^t)$} 
\label{alg:BFCOIED}
\end{algorithm}

\paragraph{Useful lemma.}
The following lemma will be useful to analyze the parameters $c$ and
$f_p$.

\begin{lemma}\label{lem:4_1}
Consider a Bloom filter with false positive rate $\frac{1}{m}$, where
$m$ is an arbitrary positive integer. Suppose at most $m$ $\sf BF.Check$ operations
are performed in the BF.
Then, for any $\delta > 0$, we have:
$$\Pr [ \mbox{\rm \# false positives} \ge 1+\delta] 
  \le {\frac {e^\delta} {(1+\delta)^{(1+\delta)}}}.$$ 
\end{lemma}

The proof, by an application of the Chernoff bound, can be found in Appendix~\ref{app:proof41}. 

Regarding the above Lemma, we remark that setting $\delta =
\Omega(\log \secparam)$, we have $$\Pr\left[\sum_{i=1}^m X_i \ge 1 +
\delta \right]  = \negl(\secparam).$$

\paragraph{Parameters $c$ and $f_p$.} We set the false positive
upperbound $f_p := \Omega(\log \secparam)$ for the BF-COIE scheme. In
our experiments, we set $f_p = 16$. 

Now, let $m = \max(2s, s+2f_p)$, we set the BF false positive rate to $1/m$.
Recall that in the BF-COIE construction, the topmost BF $B^t$ performs the $\sf
BF.Check$ operation with $2s$ times; see line (2) in
Algorithm~\ref{alg:BFCOIED}. Using the above Lemma, the number of false
positives in the top level BF $B^t$ is at most $f_p$  with all but
negligible probability in $\secparam$. Furthermore, the index $i$ in
$B^t$ is expanded into two indices $2i-1$ and $2i$ in $B^{t-1}$. This
means that the number of false indices to be checked in $B^{t-1}$ due to
the false positives in $B^t$ is at most
$2f_p$. 

Now consider an index $i$ that belongs to $B^{t}$.  Algorithm~\ref{alg:BFCOIED} will 
run ${\sf BF.Check}$ on the values $2i-1$ and $2i$ in $B^{t-1}$.  Since at least one
of these values must actually belong to $B^{t-1}$, this leads to at most one false index
being checked.
Thus, the maximum number of false indices that would be checked in
$B^{t-1}$ is at most $s + 2f_p$ (i.e., $2f_p$ from false
positives of $B^t$ and $s$ from true positives of $B^t$). 

The above argument applies inductively all the way to the bottom most
level, which means that the maximum number of false indices that would
be checked in each level BF $B^i$ will be at most $s + 2f_p$.  In the
end, the bottom BF will have at most $f_p$ false positives, and the
overall BF-COIE scheme will have at most $f_p$ false positives with all
but negligible probability in $\secparam$. 
 
For the compactness parameter $c$, we must determine the dimension $\ell$
of each BF. Recall that we set the BF false positive rate to 
$1/m$ for $m = \max(2s, s+2f_p)$:

$$ \Big(1 - e^{-\frac{\eta s}{\ell}} \Big)^\eta \le \frac 1 m.$$ 
Setting 
$\ell = \eta \cdot s \cdot m^{\frac 1 \eta}$ would
satisfy the above condition, which can be  
verified using an inequality $1 - e^{-x} \le x$ for $x \in [0,1]$; that
is, $1 - e^{-\frac{\eta s}{\ell}} \le \frac{\eta s}{\ell} =
(1/m)^{1/\eta}.$

Since the encoding has $t+1$ BFs, the overall compactness parameter is as
follows:
$$c = (t + 1) \cdot \ell = O\left ( \eta \cdot s^{1 + \frac 1 \eta}
\cdot \lg \frac n s \right ).$$

\paragraph{Efficiency.}
\BI
\item The size $c$ of encoding is $O\left ( \eta \cdot s^{1 + \frac 1 \eta} \cdot
\lg \frac n s \right )$. In our experiment, we choose $\eta = 2$. 

\item The encoding algorithm uses $O(\eta
\cdot n \cdot \lg \frac n s)$ homomorphic addition operations and hash
functions. 

\item The decoding algorithm uses $\sf BF.Check$ operations for $O(s \lg
\frac n s)$ times. 
\EI

In summary, assuming $s = o(n)$, we reduced the encoding size $c$ to be
sub-linear in $n$.  Moreover, we also reduced the number $\sf BF.Check$
operations to be sub-linear in $n$. 

\paragraph{Remark.} Although this scheme has multiple BFs, the size of
encoding $c$ is smaller than that of the warm-up scheme! This is because
with multiple levels of BFs, we can relax the false positive ratio for
each BF. The encoding computation time was increased by a multiplicative
factor of $\lg \frac n s$.

%% file: powerSums.tex
\newcommand{\ind}{\mathsf{ind}}
\subsection{COIE Scheme Based on Power Sums}
\paragraph{Removing false positives using power sums.}
We offer another encoding scheme using quite different techniques that can eliminate the false positives of the prior construction.  To achieve this,
we abandon Bloom filters, and instead use a power sum encoding, as has
been done in several works using DC-Nets for anonymous broadcast
\cite{NDSS:RufMorKat17, CCS:LYKGKM19}.

\paragraph{PS-COIE.} 
We describe a COIE scheme based on power sums, which we call PS-COIE. 
As before, we will work out the parameters after describing our
construction. The encoding algorithm is shown below. 

\begin{algorithm}[htbp]

\BEN
\item For $j = 1, \ldots, s$:
  \\ $~~~$ Compute $\lbra w_j \lket = \sum_{i=1}^n i^j \cdot \lbra v_i \lket$  
\item Output $\lbra w_1 \lket, \ldots, \lbra w_s \lket.$
\EEN
\caption{ ${\sf PS\mbox{-}COIE}.\Encode(\lbra v_1 \lket, \ldots, \lbra v_n \lket)$} 
\end{algorithm}

Note that the values of $i^j$ (modulo the underlying plaintext modulus)
are publicly computable, so computing $i^j \cdot \lbra v_i \lket$ only requires scalar multiplication and no homomorphic multiplication. 

Recall that $v_i \in \zo$. If we let $I = \{i: v_i = 1\}$ denote the
indices of the nonzero elements, then note that $$w_j = \sum_{i=1}^n i^j
\cdot v_i = \sum_{i \in I} i^j.$$ Therefore, this $w_j$ is the $j$th
power sum of the indices. Using the power sums, we present the decoding
algorithm in Algorithm~\ref{alg:pscoie}.   

\begin{algorithm}[htbp!]
\BEN
\item Recall that we have $w_j = \sum_{x \in I} x^j,$ for $j = 1,
\ldots, s$, and we would like to reconstruct all $x$'s in $I$.

\item Let $f(x) = a_s x^s + a_{s-1} x^{s-1} + \cdots + a_1 x + a_0$
denote the polynomial whose roots are the indices in $I$. 

\item Use Newton's identities to compute the coefficients of this
polynomial $f(x)$:  
\begin{align*}
a_s &= 1\\
a_{s-1} &= w_1\\
a_{s-2} &= (a_{s-1} w_1 - w_2 )/2\\
a_{s-3} &= (a_{s-2} w_1 - a_{s-1} w_2 + w_3)/3\\
\vdots\\
a_{0} &= (a_{1} w_1 - a_{2} w_2 + \cdots w_s)/s
\end{align*}
\item Extract and output the roots of the polynomial $f(x)$. 
\EEN
\caption{ ${\sf PS\mbox{-}COIE}.\Decode(\lbra w_1 \lket, \ldots, \lbra w_s \lket)$} \label{alg:pscoie}
\end{algorithm}

\paragraph{Parameters $c$ and $f_p$.} This COIE scheme has no false
positives; that is, $f_p = 0$. The compactness parameter $c$ is
equal to $s$.

\paragraph{Efficiency.}
\BI
\item The encoding algorithm uses $s \cdot n$ homomorphic addition
operations and scalar multiplications\footnote{We do not count the public multiplications to produce powers of $i$}.

\item The encoding consists of $s$ ciphertexts.

\item The decoding algorithm computes coefficients in time $O(s^2)$.
Roots of degree-$s$ polynomial can be found in time $O(s^3 \log p)$,
where $p$ is the plaintext modulus of the underlying FHE, by using the
Cantor–Zassenhaus algorithm~\cite{CanZas81}. 
\EI

%% file: countingBF.tex
\section{CODE Scheme}
In the previous section, we showed two constructions of COIE schemes for encoding a vector of indices using sublinear storage.  We now turn to the construction of CODE schemes, which, instead of encoding the indices of non-zero entries, encode the actual data values.


\paragraph{Simplified key-value store.}
To construct our CODE scheme, we first construct an auxiliary data structure that supports the
following operations: 

\BI
\item ${\sf Init}()$. Initialize the data structure.  

\item ${\sf Insert}(key, value)$. This operation allows the user to
insert an item based on its key and value.  

\item ${\sf Values}()$. Returns all  values that have been
inserted thus far. 
\EI

This data structure is simpler than a typical key-value store since it
doesn't need to find an individual item by key. Note, however, that this
is still sufficient to serve our purpose of constructing a CODE scheme. 

\subsection{BF Set}\label{ssec:countingBloom}
We now show how to instantiate a simplified key-value store using a data structure we
call a Bloom filter set ({\sf BFS}) that is in turn based on the algebraic Bloom filter
presented in Section~\ref{sec:ABF}.
To insert a pair  $(key, value)$, the Bloom
filter set stores the actual $value$ rather than an indicator bit.  Items are
inserted similar to before, by adding their value to the locations
indicated by the hashes of the $key$.

\paragraph{Input data format.}
For our construction we make an assumption on the format of the inserted data.  Specifically, we assume that all inserted values contain a unique checksum (e.g., a cryptographic hash of the value).  We assume that this checksum is sufficiently long that a random sum of checksums does not give a valid checksum except with negligible probability (as a function of $\secparam)$.


\paragraph{Construction.}
We first describe the construction of the data structure.  We show below
how to choose parameters in such a way that the client can extract all the matched items from this Bloom
filter, with
overwhelming probability.

\BI
\item ${\sf BFS.Init}() \to (B, \HHH)$.
Create an $\ell$-dimensional vector $B$ where each element can store any possible value in the domain $D$.
Choose a set of $\eta$
different hash functions  $\HHH = \{ h_q: \zo^* \to [\ell]
\}_{q=1}^\eta$. Initialize $B_i := 0$ for $i \in [\ell]$.

\item ${\sf BFS.Insert}(B, \HHH, key, \alpha)$. 
To add $(key, \alpha)$, we add $\alpha$ to the values
stored at the locations indicated by the hashes of $key$. Specifically,
  \BI
  \item For $q \in [\eta]$:
    \BI
    \item[] 
    Compute $j = h_q(key)$ and set $B_j := B_j + \alpha$. 
    \EI
  \EI

\item ${\sf BFS.Values}(B).$
  Initialize a set $V$ to be the empty set. 
  For $j \in [\ell]$, if $B_j$ has a valid checksum, add $B_j$ to $V$.
  Finally, output $V$. 
\EI

We note that, as previously proposed by Goodrich~\cite{SPAA:Goodrich11}, it is possible to avoid the checksum by maintaining a counter of the number of values inserted for each location.  Then, ${\sf BFS.Values}$ only returns values at locations with a counter of 1.


\paragraph{Parameters.}
%
We show how to set the Bloom filter parameters to guarantee that all
values can be recovered with all but negligible probability.  We assume
that we know the upper bound $s$ on the number of inserted values. We
prove the following lemma.

\begin{lemma}\label{lem:countingBloom}
If at most $s$ values have been inserted in the $\sf BFS$ data
structure, then by setting $\eta$ and $\ell$ such that $$\ell \ge 2 (s\eta - 1),$$
we can recover all $s$ values with probability at least
$1 - s\cdot(1/2)^\eta$.
\end{lemma}

\begin{proof}
Consider a (key, value) pair $(k_i,\alpha_i)$.  We say that this pair
has a {\em total collision} if every hash position for the pair is also
occupied by another inserted key, value pair.  In this case, $\alpha_i$
cannot be recovered. On the other hand, if at least one hash position
has no collisions, then we can recover the value. Note that the
collision depends on the key $k_i$ but not the value $\alpha_i$. 

For a given key $k_i$, we define the event \textsf{TCOL}$(k_i)$: $${\sf
TCOL}(k_i) = 1 \textrm{ if } \forall q \in [\eta], \exists (k', q')
\ne (k_i,q): h_q(k_i) = h_{q'}(k').$$
Here, $k'$ can be the key of any item that has been inserted in the
set. Since the set contains at most $s$ items, there are at most $s$
possible keys for $k'$. Recall also that $\eta$ hash functions are applied
for each item. 

Since for each $k_i$, there are at most $\eta s - 1$ pairs of $(k',
q')$s that are different from $(k_i,q)$, we can bound the collision
probability as follows:
\[ \Pr[{\sf TCOL}(k_i) ] \le \left(\frac{(\eta s-1)}{\ell}\right)^\eta \]

Thus, if we choose $\eta$ and $\ell$ such that $\ell \ge 2 (s\eta - 1)$,
we have
\[ \Pr[{\sf TCOL}(k_i) ] \le (1/2)^\eta\]

Taking a union bound over all $s$ inserted values, we have
\[\Pr[ \exists k_i : {\sf TCOL}(k_i) ] \le s \cdot (1/2)^\eta\].
\end{proof}

\subsection{CODE Scheme Based on BF Set}
In this section, we construct a CODE scheme. Recall that unlike encoding
the indices through a COIE scheme, a CODE scheme encodes data in a
compressed manner. The main idea of our construction is simulating the
operations of ${\sf BFS}$; we call our scheme $\sf BFS\mbox{-}CODE$.


\paragraph{Pre-processing the input data.}
As mentioned in the description of the BF Set construction, we need to pre-process the input data so that each item is
attached with its checksum. Although a data item $v$ is represented as a
single number, it is assumed that $v$ can be parsed as $v.val$ for its
actual value and $v.tag$ for its checksum. 
Moreover, we assume that the checksum is long enough, such that a random
linear combination of checksums is only negligibly likely to produce a
valid checksum (i.e., $|checksum| = \omega(\lambda)$). 

We stress that when our CODE scheme is used for secure search, this pre-processing can be performed locally 
by the client prior to encrypting his data.  Moreover, computing
checksum adds only a tiny amount of overhead. 

\paragraph{BFS-CODE.}
We now describe our $(n, s, c, f_p)$-BFS-CODE construction over domain $D$. As before,
we will work out the parameters after describing our construction. The
encoding algorithm is shown below. 

\begin{algorithm}[htbp]
\BEN
\item $\eta = \secparam+\lg{s}$; $\ell = 2 (\eta s - 1)$
\item  Initialize $\lbra B \lket = (\lbra B_1 \lket, \ldots, \lbra
B_\ell \lket) := (\lbra 0 \lket, \ldots, \lbra 0 \lket)$. 
\item Choose $\HHH = \{ h_q: \zo^* \to [\ell] \}_{q=1}^\eta$ at random.
\item For $i \in [n]$ and for $q \in [\eta]$:
    \BEN
        \item[] $j := h_q(i)$; 
                $\lbra B_j \lket = \lbra B_j \lket + \lbra v_i \lket$
    \EEN
\item Output $\lbra B \lket$.
\EEN

\caption{ ${\sf BFS\mbox{-}CODE}.\Encode(\lbra v_1 \lket, \ldots, \lbra v_n \lket)$} 
\end{algorithm}

Note that at step 4 in the above, if $v_i$ is 0, then $B_j$ stays the
same. On the other hand, if $v_i$ is not 0, $B_j$ will be increased by
$v_i$.  This implies that $B$ will exactly hold the result of 
operations $\{{\sf BFS.Insert}(B, \HHH, i, v_i): i \in \idx(v) \}.$

The decoding algorithm is simple, and it's described in
Algorithm~\ref{alg:BFSD}.
\begin{algorithm}[htbp]
\BEN
\item Output ${\sf BFS.Values}(B)$
\EEN
\caption{ ${\sf BFS\mbox{-}CODE}.\Decode(B)$} 
\label{alg:BFSD}
\end{algorithm}

\paragraph{Correctness.} This is immediate from the additive
homomorphism of the underlying encryption scheme and the parameters for
the $\sf BFS$. In particular, we set $\eta = \secparam+\lg{s}$ so that the
probability of recovery error is at most $2^\secparam$. 

\paragraph{Parameters $c$ and $f_p$.}
The checksums attached to the data items ensure that we have no false
positives with overwhelming probability, that is, $f_p = 0$. The compactness
parameter $c$ is the dimension $\ell$ of the BF, which is $O(\eta s)$. 

\paragraph{Efficiency.}
\BI
\item The encoding algorithm uses $\ell = O(\eta s)$ encryption
operations, $\eta \cdot n$. 
addition operations, and $\eta n$ hash functions. 
\item The encoding consists of $\ell$ ciphertexts. 
\item The decoding algorithm uses $\ell$ decryption operations.
\EI

Since by Lemma~\ref{lem:countingBloom}, the size $\ell$ of the Bloom
filter only depends on the number of matches $s$ and the number of hash
function $\eta$, we get that the communication complexity of the above
protocol is independent of the database size $n$.



%% file: ssp.tex
\section{Secure Search Protocols}
\label{sec:ssp}
We implement secure search protocols by using compressed oblivious
encoding schemes. We begin by defining a relaxed notion of correctness
that allows for false positives, as is needed in some of our
constructions.  we then define security of secure search.

\subsection{$(\ell, f_p)$-Relaxed Secure Search} 

We relax the correctness guarantee to allow the Client to retrieve  a
superset of the matching records.  Specifically, if $\mathcal{S}$ is the
set of indexes matching a Client's query $q$, then at the end of the
protocol, we require the Client to obtain a set $\mathcal{S}'$ such
that:
\begin{itemize}
    \item With all but negligible probability, $\mathcal{S} \subseteq \mathcal{S}'$
    \item With all but negligible probability, $|\mathcal{S}'\setminus \mathcal{S}| \leq f_p$.
\end{itemize}

We parameterize a secure search scheme by $(\ell, f_p)$, where $\ell$ is
the \emph{amortized} communication complexity {\em per} matching record, and $f_p$ is the
number of ``false positives,'' as defined above.

\subsection{Security of Setup-free Secure Search} 
To define security of our secure search schemes, we use a game-based
security definition similar to that of Akavia et al.~\cite{PETS:AGHL19}.
The game is between a challenger and an adversary $\A$ with regard to a
setup-free search scheme, $\ssp$, and an FHE scheme, $\fhe$.


\begin{enumerate}
\item[] \hspace{-2em} $\Game$:

\item The challenger runs a key generation algorithm (with computational security
parameter $\kappa$) and sends the evaluation key to $\mc{A}$ so that $\mc{A}$
can perform homomorphic additions and multiplications. 

\item $\A$ chooses either:
    \begin{itemize}
        \item Two databases $x^0 = (x^0_1, \ldots, x^0_n)$ and $x^1 =
        (x^1_1, \ldots, x^1_n)$ of the same length, and a query $q$, or  
        \item A single database $x = (x_1, \ldots, x_n)$ and two queries
        $q^0, q^1$ of the same circuit size. 
    \end{itemize}
    In both cases, we require that the sizes of the two result sets (denoted by $s$) are equal.
    
    \item The challenger samples $b \leftarrow \{0,1\}$.  Then, either 
    \begin{itemize}
        \item Runs Setup on input $x^b$ and the search protocol from $\ssp$ on input $q$, or
        \item Runs Setup on input $x$, and the search protocol from
        $\ssp$ on input $q^b$.
    \end{itemize}
    \item $\A$ outputs a bit $b'$
    \item We say that $\A$ has advantage $$\Adv = |\Pr[b=b']-1/2|.$$
\end{enumerate}

\begin{definition}
A setup-free $(\ell, f_p)$-secure search scheme $\ssp$ is \emph{fully
secure} if every PPT adversary $\A$ controlling the server has a
negligible advantage $\Adv \le \negl(\kappa)$ in the game above.
\end{definition}

\subsection{From COIE to Secure Search}

We next present our framework for obtaining Secure Search from COIE.  
The intuition is likely already clear from the previous descriptions:~the encrypted client query is applied 
to the dataset, returning an encrypted bit vector indicating where index matches lie.  The server homomorphically computes the 
hamming weight of this vector, and sends it to the client for decryption. This provides the result set size to the Server, allowing it to encode the result vector in the COIE.\footnote{We note if we don't wish to reveal this to the server, we can use a fixed, global upper bound, or, if it is appropriate to the application, the client can add noise to provide differential privacy.  It is also worth pointing out that prior work leaks the result set size as well.}  The encoding is sent to the client for decryption and decoding.  

Because the COIE only encodes the indices, and not the data values, we 
then add a PIR step to fetch the corresponding data.  Note that if 
the COIE scheme admits false positives, it is possible that the number of false positives, and therefore the number of PIR queries, depends on the data, 
leaking something to the Server.  To fix this problem, the client pads the number of PIR queries as follows.  It 
fixes a bound $f_p$ on the number of false positives, and aborts if the actual number of false positives exceeds this bound.  Otherwise, the client uses enough dummy queries to pad the number of PIR queries to $s+ f_p$. 

\begin{algorithm}[htbp]
\small

\begin{enumerate}
\item Client runs the FHE key generation algorithm and encrypts database
$x = (x_1, \ldots, x_n)$ with $x_i \in \zo^m$. It then sends $\lbra x
\lket = (\lbra x_1 \lket, \ldots, \lbra x_n \lket)$ and the evaluation
key to Server.

\item Client sends an encrypted query $\lbra q \lket$.

\item Server homomorphically evaluates the encrypted query $\lbra q
\lket$ on each encrypted record. In particular, let $\lbra b \lket =
(\lbra b_1 \lket, \ldots, \lbra b_n \lket)$ where $\lbra b_i \lket =
\lbra q(x_i) \lket$.  Note that $q(x_i) = 1$ if record $i$ is a
match and is equal to $0$ otherwise.

\item Server homomorphically computes $\lbra s \lket = \sum_{i=1}^n \lbra b_i \lket$, and sends to Client for decryption. 

\item Client decrypts $\lbra s \lket$ to obtain $s$, and sends $s$ to Server. 

\item Server calls \textsf{COIE}.$\Encode(\lbra b \lket)$ with sparsity parameter $s$, to obtain an encrypted
encoding $\lbra C \lket$.  It sends $\lbra C \lket$ to Client. 

\item  Client decrypts
$\lbra C \lket$ into $C$ and calls \textsf{COIE}.$\Decode(C)$ to obtain a set $\mathcal{S}'$
of size $s + e$ indexes.  If $e > f_p$, Client aborts. Otherwise,
Client adds $f_p -e$ number of dummy indexes to $\mathcal{S}'$.

\item Client runs a PIR protocol with the Server to obtain the records
corresponding to the indexes in $\mathcal{S}'$.

\end{enumerate}
\caption{\small Secure search with a $(n, s, c, f_p)$-COIE scheme.}\label{alg:ssCOIE}
\end{algorithm}

\begin{theorem}
Given an FHE scheme, a $(n, s, c, f_p)$-COIE scheme in the random oracle model, and a PIR scheme in the random oracle model with
communication complexity $\ell_p$ for records in $\zo^m$, the construction in Algorithm \ref{alg:ssCOIE} yields a $(\ell,f_p)$-secure search scheme for records in $\zo^m$ in the Random Oracle
Model, where $\ell = \frac{c\cdot \ell_c + (s+f_p)\cdot \ell_{p}}{s}$,
$\ell_c$ is the length of an FHE ciphertext, and $s$ is the number of
matching records.  
\end{theorem}

\begin{proof}
We begin by proving that the adversary cannot distinguish between two
different queries.  The adversary chooses a database $x$ and two
queries $q^0$ and $q^1$, with the promise that $s = \sum_{i=1}^n q^0(x_i) =
\sum_{i=1}^n q^1(x_i)$.

The entire view of the adversary during the experiment can be
reconstructed efficiently given (1) the encrypted database $\lbra x
\lket$ (2) the encrypted query $\lbra q \lket$, (3) $s + f_p$ iterations
of the PIR protocol, requesting indexes in $\mathcal{S}'_b$, where $s$
is the number of matching records.

Since the value of $s$ is the same for $q^0$ and $q^1$, the two things
that change in the view of the adversary when switching from $b = 0$ to
$b=1$ are (1) the encrypted query $\lbra q^b \lket$ (2) the set of
indexes $\mathcal{S}'_b$ (but not the number) requested during the PIR
step.  

We also note that the experiment only aborts when the number of received false positives $e$ is greater than the bound $f_p$, which only happened with probability $\negl(\secparam)$ for a statistical security parameter $\secparam$.  Thus, we ignore this possibility in the following.  

We can now proceed via a standard hybrid argument: 
\BI
\item We first consider the real experiment with $b=0$. 

\item We then switch the encrypted query from $q^0$ to $q^1$, but leave
the set of indexes in the PIR step as $\mathcal{S}'_0$.
Indistinguishability of the adversary's view follows from the IND-CPA
security of the FHE scheme.  

\item Next, we switch the set of indexes in the PIR step from
$\mathcal{S}'_0$ to $\mathcal{S}'_1$.  Indistinguishability of the
adversary's view now follows from the security of the PIR scheme. This
is now identical to the real experiment with $b=1$.  
\EI

We conclude that the probability the adversary outputs $0$ or $1$
differs by a negligible amount when $b=0$ versus $b=1$.  Therefore, the
advantage of the adversary in guessing $b$ is negligible.

The proof that the adversary cannot distinguish between the same query
applied to two different databases follows nearly identically.
\end{proof}

\subsection{From CODE to Secure Search}

We next present our framework for obtaining
Secure Search from CODE.

\begin{algorithm}[htbp]
\small

\begin{enumerate}
\item Client runs the FHE key generation algorithm and encrypts database
$x = (x_1, \ldots, x_n)$ with $x_i \in D$. It then sends $\lbra x
\lket = (\lbra x_1 \lket, \ldots, \lbra x_n \lket)$ and the evaluation
key to Server.

\item Client sends an encrypted query $\lbra q \lket$.

\item Server homomorphically evaluates the encrypted query $\lbra q
\lket$ on each encrypted record. In particular, let $\lbra b \lket =
(\lbra b_i \lket, \ldots, \lbra b_n \lket)$ where $\lbra b_i \lket =
\lbra q(x_i) \lket$.  Note that $q(x_i) = 1$ if record $i$ is a
match and is equal to $0$ otherwise.

\item Server homomorphically computes $\lbra s \lket = \sum_{i=1}^n \lbra b_i \lket$
 and sends $\lbra s \lket$ to Client. 

\item Client decrypts $\lbra s \lket$ to obtain $s$ and sends it back
to the Server.
  
\item Server computes $\lbra d_i \lket = \lbra b_i \lket \cdot \lbra x_i
\lket$ for $i \in [n]$. Then, it applies \textsf{CODE}.$\Encode(\lbra d_1 \lket,
\ldots \lbra d_n \lket)$ with sparsity parameter $s$, to obtain an encrypted encoding $\lbra C
\lket$.  It sends $\lbra C \lket$ to Client. 

\item Client decrypts $\lbra C \lket$ to $C$ and decodes $C$ to obtain a
set $\mathcal{S}$ of size $s$ matching records.

\end{enumerate}
\caption{\small Secure search with a $(n, s, c, f_p)$-CODE scheme.}\label{alg:ssCODE}
\end{algorithm}

\begin{theorem} \label{thm:ssp-code}
Given an FHE scheme, and a $(n, s, c, f_p)$-CODE scheme over domain $D$
in the random oracle model, the construction in Algorithm~\ref{alg:ssCODE} yields a $(\ell,
f_p)$-secure search scheme for records in domain $D$ in the random oracle
model, where $\ell = \frac{c(s)\cdot \ell_c}{s}$, $\ell_c$ is the length
of an FHE ciphertext with plaintext space $D$, and $s$ is the number of
matching records.
\end{theorem}

The proof is similar to the COIE-based scheme and can be found in
Appendix~\ref{app:proof63}. 

\paragraph{On the use of homomorphic multiplication.}
As described, our CODE-based search scheme uses $n$ homomorphic
multiplications to create the vector $\lbra d \lket$.  However, it may
be the case that this vector is already produced as part of the match
step, for example for arithmetic queries.  In this case, our CODE scheme
requires no further homomorphic multiplications.

\paragraph{On volume attacks.}
In our secure search schemes, the client sends the number $s$ of matching
records to the server so that the server can create an oblivious compress
encoding. 
One recent line of works has developed attacks using volume leakage (e.g.,
\cite{CCS:KKNO16,CCS:GuiJohWar19,NDSS:BlaKamMoa20}), and these types of attacks
can be applied to our scheme in theory. 

In our scheme, the volume attacks can be mitigated by hiding $s$ in
a differentially private manner. In particular, the client can add a small
amount of noise to $s$ before sending it to the server. A similar approach was
used in previous work e.g.,~\cite{CCS:PPYY19}.

%% file: exp.tex
\section{Evaluation}

\subsection{Fetch time}
We implemented our search protocols based on BF-COIE, PS-COIE, and
BFS-CODE schemes. 
All protocols were implemented using PySEAL~\cite{pyseal}, which
is a Python wrapper of the Microsoft research SEAL library (version
3.6)~\cite{sealcrypto} using the BFV encryption scheme~\cite{BFV}.  We
instantiated a single-server PIR protocol in our construction using
SealPIR~\cite{SP:ACLS18}. For the root finding step of the decoding
procedure in PS-COIE, we use an implementation based on SageMath
9.2~\cite{Sage}.
 
\paragraph{Measuring the Fetch step.}
Our search framework improves the overall search time by executing the
Match step only once, while the LEAF protocol must execute the Match
step $s$ times. However, since we do not optimize the Match step itself
over prior work, we focus on measuring the cost of the Fetch procedure.
That is, our experiments measure the time from when the server holds
encrypted query results, i.e., $(\lbra b_1 \lket, \ldots, \lbra b_n
\lket)$ with $b_i \in \zo$, to when the client recovers all $s$ records
matching the query.  Specifically, we measure the cost of steps 4 and up
in Algorithms~\ref{alg:ssCOIE} and~\ref{alg:ssCODE}.  Similarly, for
LEAF+, we only measure the cost of the Fetch step.  

\paragraph{Database.}
To measure the performance of our protocols, we run experiments with
database size $n$ ranging from 1000 to 100,000 data items and the result
set size $s$ set to between 8 and 128.  As in the LEAF+
experiments~\cite{CCS:WYXZ20}, all data items are $16$-bit integers. 

\paragraph{BF-COIE parameters.}
For the BF-COIE secure search, we set the parameters as indicated in
Section~\ref{sec:BF-COIE}. 
\BI
\item 
We set the false positive upperbound $f_p = 16$.  Recall that the client
aborts (without executing the PIR) if the actual number of false
positives exceeds this, but this only happens with probability
negligible in the security parameter, which we set $\secparam = 40$.

\item We set the number of hash function $\eta=2$ for each Bloom filter, so
  each BF has size $\ell = 2s \cdot \sqrt{2s}$. (If $2s < s + 2f_p$, we set
  $\ell = 2s \cdot \sqrt{s + 2f_p}$). 
\EI

\paragraph{BFS-CODE parameters.}
For BFS-CODE secure search, with $\secparam = 40$, the number of
hash functions $\eta$ is set to $\secparam+\lg{s}$, and the Bloom filter
size is set to $2(\eta s - 1)$. 
Additionally, each data item is attached with a 40-bit checksum to
guarantee a $2^{-\secparam}$ probability of collision. We used SHA2 to
compute a checksum.   

\paragraph{Implementing LEAF+.}
For a comparison we also implemented the fetch step of the LEAF+
protocol~\cite{CCS:WYXZ20}, since their implementation is not publicly available.

Their protocol has $O(\log \log n)$ depth of multiplications. Therefore, they
have to use bootstrapping techniques to reduce the accumulated noise.
However, SEAL doesn't provide a method for bootstrapping, and we suspect
that they added a customized implementation of bootstrapping on top of
SEAL. Unfortunately, their implementation is not available. 

We address this issue by choosing to ignore the time for bootstrapping
when we measure the running time of our implementation of LEAF+. Of
course, our implementation doesn't output the correct results, but the
measured running time will be shorter than the actual running time.
Therefore, we believe that this measured time serves as a good baseline.  

\paragraph{Experiment environments.}
All our experiments were performed on an Intel\textregistered Core 9900k
@4.7GHz with 64GB of memory. For fair comparison, the test was performed
on a single thread with no batching optimizations for computation.
Networking protocol between server and clients is a 1Gbps LAN.


\begin{figure}
\includegraphics[width=.95\linewidth]{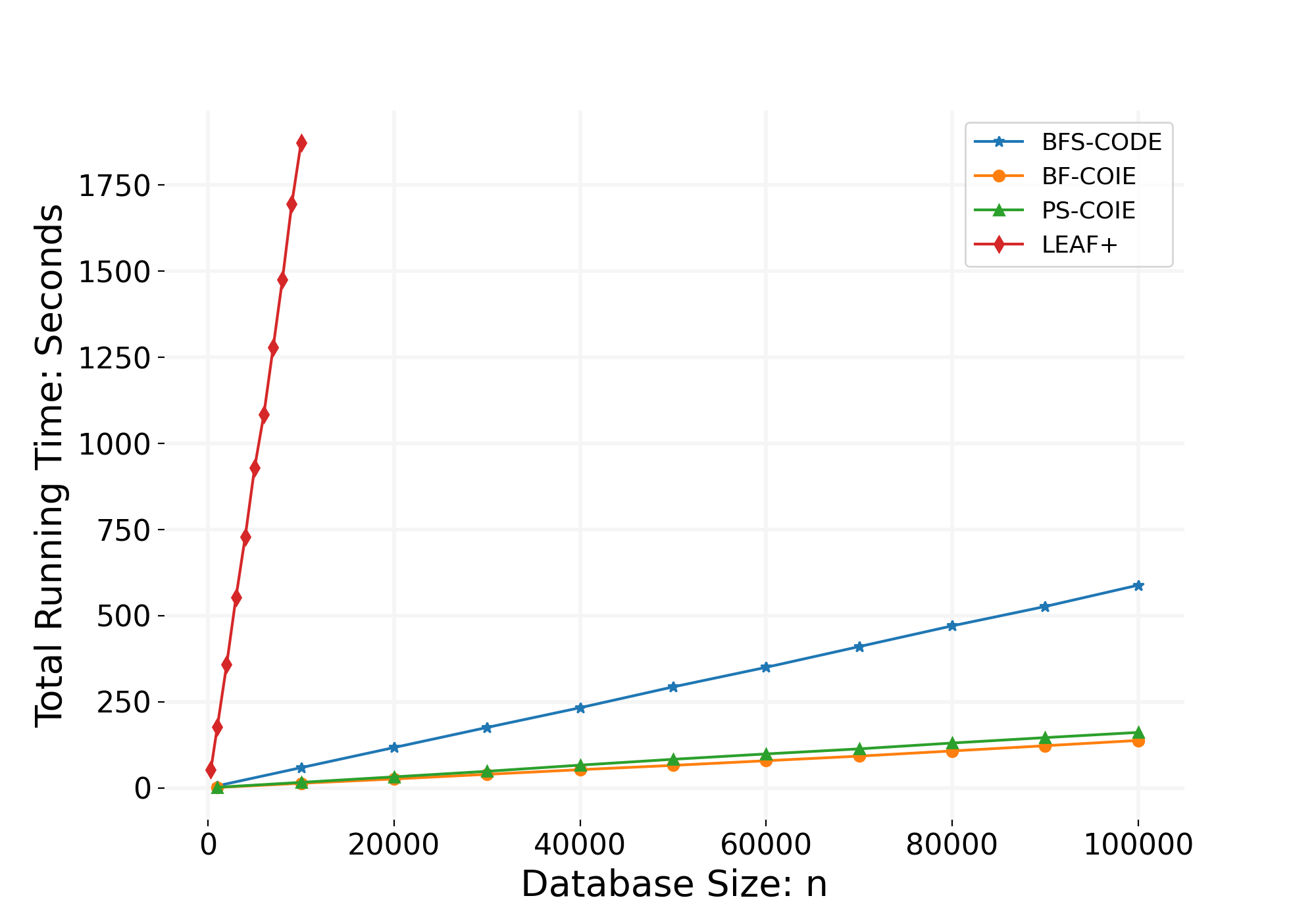}
\BI
\item[]
{\small For {LEAF+}, we plot the time for fetching only a {\bf
single} record, since fetching $s$ records takes too long.}
\EI
\caption{Fetch time vs. Database size with $s=16$.}
\label{fig:t_n}
\end{figure}

\paragraph{Results: Fetch time vs.~database size.}
Figure~\ref{fig:t_n} shows the performance of our protocols as a
function of database size, while the result set size $s$ is fixed to 16.
However, for {LEAF+}, we plot the time for fetching only a {\bf single}
record, since fetching $s$ records takes too long. 
In our implementation of LEAF+, fetching even a single record when $n=10,000$ requires
1872 seconds. We note that the authors of LEAF+ report about 60
seconds for a single fetch~\cite{CCS:WYXZ20}. We conjecture that they parallelize the scheme with 32 threads.
Here, we only use a single thread. 

\begin{figure}
\includegraphics[width=.95\linewidth]{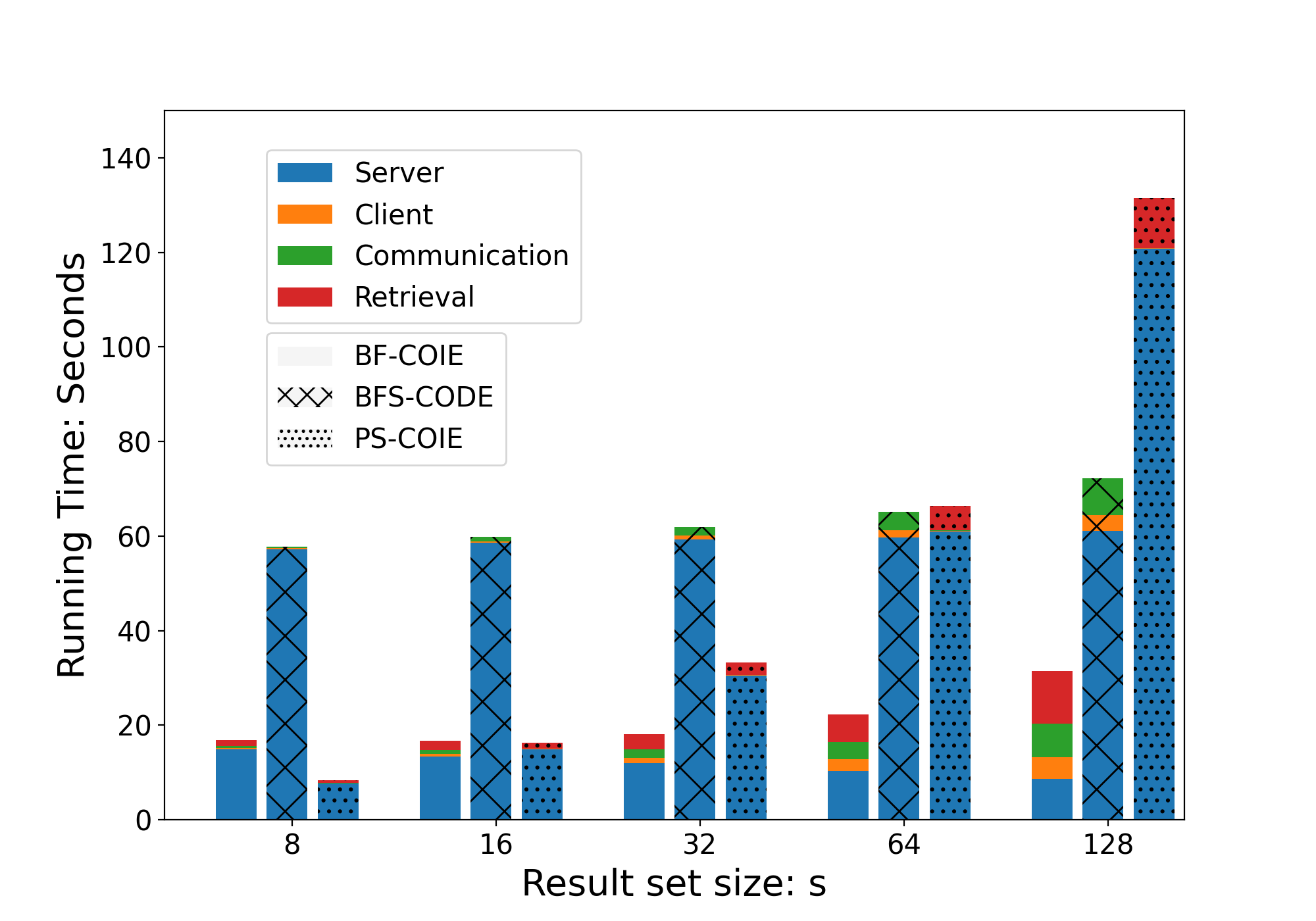}
\vspace{-1em}
\caption{Fetch time vs.~Result set size with $n=10,000$.}
\label{fig:ct_s}
\vspace{-1em}
\end{figure}

All three of our protocols greatly outperform LEAF+. Looking at BF-COIE in
particular:
\BI
\item In BF-COIE search, fetching 16 records with $n=10,000$ takes 16.7 seconds,
  compared to 1872 seconds for a single record fetch in LEAF+.  We believe that
  the speed up is due to the fact that LEAF+ (with a single-record fetching)
  needs $O(n \log n)$ homomorphic additions and $O(n)$ homomorphic
  multiplications, while BF-COIE search needs only $O(n \log \frac n s)$
  homomorphic additions with {\em no homomorphic multiplications.}  In
  addition, as Figure~\ref{fig:ct_s} shows, the overhead of the PIR step to
  retrieve the actual data is small. 

\item 
Due to the sequential limitation in LEAF+, fetching $16$ records with
LEAF+ is extrapolated to take about $16 \cdot 1872 = 29952$ seconds.
Overall, BF-COIE search is about {\bf 1800 times faster} than LEAF+.
\EI

The time for all three of our protocols is dominated by the server's
computation during encode, which grows linearly with the DB size.  

Since the number of hash functions $\eta$ is larger in the BFS-CODE
protocol than in BF-COIE protocol, the encoding step of this protocol
takes longer.  


\paragraph{Results: fetch time vs.~the result set size.}
Figure~\ref{fig:ct_s} shows the performance of our protocols as a
function of the result set size $s$ while $n$ is fixed to $10,000$.
Here, again the performance is dominated by the encoding step, but the
relative costs have changed.  Due to the need to compute more power
sums, the PS-COIE protocol performs worse than BS-COIE and BFS-CODE when
$s$ becomes moderately large.  

The time used for transmitting the data over network (green in
Figure~\ref{fig:ct_s}) increases for larger $s$. However, it still
remains small for all three schemes.
In the scenario of having lower network bandwidth, batching is
recommended to pack a vector of ciphertexts into a single ciphertext
with relatively low computation overhead. We discuss communication costs
further in Section~\ref{sec:comm}. 

\subsection{Overall Running Time}

Although we do not optimize the Match step itself over prior work, we provide
an estimated comparison of the running time for the end-to-end flow.

Our search framework improves the overall search time by executing the Match
step only once, while the LEAF protocol must execute the Match step $s$ times.
Based on this, we can extrapolate the running time as follows: 

\BI
\item The overall running time for LEAF: 
$$ Time({\sf LEAF}) = s \cdot {\sf MT}({\sf LEAF})
  + s \cdot {\sf FT}({\sf LEAF}).$$ Here, $\sf MT$ and $\sf FT$ denote the
  match time and fetch time respectively. 

\item The overall running time for the BF-COIE scheme: $$Time({\sf
BF\mbox{-}COIE}) = {\sf MT}({\sf BF\mbox{-}COIE})
  + {\sf FT}({\sf BF\mbox{-}COIE})$$
\EI

Although the implementation (nor the algorithm) of the matching step of
LEAF protocol is not available in~\cite{CCS:WYXZ20}, we expect that it
holds 
${\sf MT}({\sf LEAF}) \approx {\sf MT}({\sf BF\mbox{-}COIE})$. 
In the experiment performed in LEAF (see Figure 9 in~\cite{CCS:WYXZ20}),
we have $m = \frac{ {\sf MT}({\sf LEAF})}{{\sf FT}({\sf LEAF})} \approx
1.5$.
For $s=16$, setting $ {\sf FT}({\sf LEAF}) = 1800 \cdot {\sf FT}({\sf
BF\mbox{-}COIE})$ based on the above discussion, we can estimate the
speed-up as follows:
$$ \frac {Time({\sf LEAF})}{Time({\sf BF\mbox{-}COIE})} = \frac{ s \cdot (m + 1) } { m + 1/1800}.$$ 
Thus, with $s = 16$, we estimate that our BF-COIE scheme has roughly 26X end-to-end speed-up.

\subsection{Communication} \label{sec:comm}
We now look at the communication required by each of our schemes and by LEAF+.
Figure~\ref{tbl:communication_new} shows the network cost of the
protocols when the result set size $s$ is 16 and the size of the
database is $n=10,000$.  
In our implementations, the length of an FHE ciphertext is
approximately 103KB and the communication cost of PIR is
approximately 369KB. 

\begin{figure}[h]
\small
\begin{tabular}{lcccc}
\hline 
& LEAF+ & BF-COIE & PS-COIE & BFS-CODE \\
\hline
\#ct's  & $704$ & $1323$ & $17$ & $1321$ \\
\#PIR & 0 & 32 & 16 & 0 \\
\hline
\#ct's (w/~batching) & $32$ & $2$ & $2$ & $2$ \\
\hline
\end{tabular}

\caption{The communication costs ($n=10,000$ and $s=16$).} 
\label{tbl:communication_new}
\end{figure}

To explain this table, we first need to explain how we determined the costs of LEAF+ and PIR.

\BI
\item {\it LEAF+.}
Since LEAF+ fetches each data item and the corresponding index one by
one, LEAF+ needs to 16 rounds of communication to retrieve 16 data
items. Worse yet, LEAF+ requires the client to send the index of the
previous match (requiring $\lg{n}$ bits) in his next query to ensure
correctness.  Finally, LEAF+ uses bitwise encryption requiring a
ciphertext for each bit of the encrypted communication.  Thus, in a
single round, the client must send $\lg n=14$ ciphertexts and the server
returns $16+\lg n = 30$ ciphertexts -- $16$ ciphertexts for returning
the matching data item, and $\lg n$ ciphertexts to return its index.
This amounts to 704 ciphertexts for fetching 16 items (excluding the
query).    

\item {\it PIR costs.}
We reduce the cost of PIR for the COIE-based schemes by making a slight modification.
In addition to storing the FHE-encrypted database, the server also stores a copy
of each record encrypted using a symmetric-key encryption scheme (resulting in much
shorter ciphertexts).  Then, in the PIR step, the client fetches this symmetrically encrypted ciphertext 
instead of the FHE-encrypted one.

We use SealPIR for our PIR protocol, which requires 368.6 KB per request.
We remark that a very recently introduced SealPIR+ takes 80KB per
request~(see Table~1 in~\cite{MulPIR}), using which we can reduce the
communication further.
\EI

We can now compare the communication costs based on rows 1 and 2 of
Figure~\ref{tbl:communication_new}.  We see that the communication of
BF-COIE and BFS-CODE are roughly twice that of LEAF+, while PS-COIE
requires almost 10X less communication.  The extra communication needed
by BF-COIE and BFS-CODE can likely be offset by the much lower round
complexity required by our protocol since the latency costs are likely
higher than the cost for the extra bandwidth.

\paragraph{Reducing communication using ciphertext batching.}
We now describe an optimization to significantly reduce the
communication of our protocols at the cost of slightly increased server
computation.  SEAL allows thousands of encrypted values to be packed
together into a single ciphertext.  This allows us to pack the
ciphertexts in all of our protocols into just one a single ciphertext to
be sent from the server to the client.  However, this does require the
server to do some additional computation to pack the ciphertexts prior
to sending them.  We experimentally measured this packing, and it requires
approximately 3 seconds on a single threaded machine.

LEAF+ can also take advantage of packing to reduce the communication of
their protocols.  However, since the results must be returned one at a
time, the best LEAF+ can do is to pack all ciphertexts that are sent in
each round, resulting in a total of 32 ciphertexts. 

We note that the cost of PIR is unchanged by this modification.  Thus,
with the packing optimization, the communication of BFS-CODE is roughly
1/16 of the communication needed by LEAF+, but BF-COIE and PS-COIE
require approximately 4X and 2X more communication than LEAF+
respectively when SealPIR is used; however, when SealPIR+ is used, both
schemes have slightly less communication than LEAF+.

%% file: related.tex
\section{Related Work}

\subsection{Techniques for Secure Search}
\paragraph{Secure pattern matching (SPM) on FHE-encrypted data.} 
In SPM, given an encrypted query $\lbra q \lket$ and $n$
FHE-encrypted data items $(\lbra x_1 \lket, \ldots, \lbra x_n \lket)$, it
returns a vector of $n$ ciphertexts $\lbra b_1 \lket, \ldots, \lbra b_n
\lket$, where $b_i$ indicates whether the $i$th data element is a
match~\cite{CCSW:YSKYK13,FCW:CheKimLau15,TIFS:CKK16,TDSC:KLLTW19}.
Their works focus on optimizing the search circuits to determine whether a data item matches the query, and therefore
the communication complexity and client’s running time are proportional
to the number of data items. Our work focuses on the orthogonal
problem of optimizing the retrieval of the matched data items with
sublinear communication and client computation.  

\paragraph{Searchable encryption (SE).} 
Searchable encryption~\cite{SP:SonWagPer00,EC:BDOP04} allows highly efficient search (usually in
$o(n)$ time) over encrypted data.  Efficient SE schemes have been proposed for a wide variety of queries including equality queries~\cite{CCS:CGKO06, AC:ChaKam10}, range queries~\cite{RSA:IKLO16, CCS:RACY16}, and conjunctive queries~\cite{SP:PKVKMC14, C:CJJKRS13}. However, to achieve sublinear query performance, SE schemes require significant preprocessing and relax security, allowing some partial information about the queries and data (e.g. access patterns) to leak to the server.  For a recent survey on SE constructions and security, see Fuller et al.~\cite{SP:FVYSHG17}.
In contrast, our work focuses on achieving preprocessing-free secure constructions, leaking nothing about the queries or results other than their sizes. 

\paragraph{Property Preserving Encryption (PPE).}
As a different approach, property-preserving encryption~\cite{EC:PanRou12} produces ciphertexts that maintain certain relationships (e.g., equality, and order) of the underlying plaintexts.  This allows queries to be performed over ciphertexts in the same way that they can be carried out over plaintexts.  Examples of PPE include deterministic encryption~\cite{C:BelBolONe07} allowing equality queries, and order-preserving encryption~\cite{EC:BCLO09, C:BolCheONe11} allowing range queries.  However, it has been shown~\cite{NDSS:IslKuzKan12, CCS:GMNRS16,SP:GSBNR17} that such property-preserving ciphertexts leak a lot of information about the underlying plaintexts.
See~\cite{SP:FVYSHG17} for a survey of constructions and attacks.

\subsection{General Techniques}
\paragraph{Private information retrieval (PIR).} PIR
allows the client to choose the index $i$ and retrieve the $i$th record
from an untrusted server while hiding the index
$i$~\cite{JACM:CKGS98}. However, this protocol by itself provides only a
limited search functionality requiring the client to know the index of the data to retrieve.  In this work, we
aim at protocols supporting any arbitrary search
functionality. 

\paragraph{Secure multi-party computation (MPC).}
Secure two-party computation~\cite{FOCS:Yao86,STOC:GolMicWig87} allows
players to compute any function of their private inputs without
compromising privacy of their inputs. For example, the client and the
server can run a protocol for secure two-party computation to solve the
secure search problem. 
While there has been much progress in improving efficiency of MPC
protocols, such protocols still require $\Omega(n)$ communication and $\Omega(n)$ client computation per query.  In this work, we aim to achieve protocols with sublinear communication and client work.



\paragraph{Oblivious RAM (ORAM) and Oblivious data structure (ODS).} 
ORAM~\cite{GO96} is a protocol which allows a client to store an array
of $n$ items on an untrusted server and to access an item obliviously,
that is, hiding contents and which item is accessed (i.e., the access
pattern).  Likewise, ODS~\cite{CCS:WNLCSS14} allows the client to store
and use a data structure obliviously. 
One could implement secure search by utilizing an ODS for a search tree.
However, ODS constructions typically need $\Omega(\log^2 n)$ rounds for
each operation. In this work, we aim at achieving a constant round
protocol.

%% file: conclusion.tex
\section{Conclusion}
We have presented several new constructions of secure search based on fully homomorphic encryption.  Prior constructions were
inherently sequential, returning only a single record from the result set, and requiring a new query from the client that depended on the index of the previous match.  We have demonstrated several new methods for 
encoding the entire result set at one time, 
removing the added rounds, and allowing the
server work to be parallelized.  
Additionally, we have shown that this can be done without homomorphic multiplication, 
ensuring low computational cost at the server. Finally, we have implemented our constructions, and demonstrated up to three orders of magnitude speed-up over prior work.  Additionally, we introduced the notion of compressed oblivious encoding which may be of independent interest.

%% file: Acknowledge.tex
\section*{Acknowledgements}
Dana Dachman-Soled is supported in part by NSF grants CNS-1933033, CNS-1453045(CAREER), and by financial assistance awards 70NANB15H328 and 70NANB19H126 from the U.S. Department of Commerce, National Institute of Standards and Technology; 
Seung Geol Choi is supported by ONR N0014-20-1-2745 and NSF grant CNS-1955319; 
S. Dov Gordon is supported by the NSF Grants CNS-1942575 and CNS-1955264, by the Defense Advanced Research Projects Agency (DARPA) and Space and Naval Warfare Systems Center, Pacific (SSC Pacific) under Contract No. N66001-15-C-4070, by the Blavatnik Interdisciplinary Cyber Research Center at Tel-Aviv University and
Israel National Cyber Directorate (INCD), and by a Google faculty award; 
Arkady Yerukhimovich is supported by NSF grant CNS-1955620, and by a Facebook Research Award.  

%% file: app.tex
\appendix

\section{Proof of Lemma~\ref{lem:4_1}}\label{app:proof41}

\begin{lemma}[4.1]
Consider a Bloom filter with false positive rate $\frac{1}{m}$, where
$m$ is an arbitrary positive integer. Suppose at most $m$ $\sf BF.Check$ operations
are performed in the BF.
Then, for any $\delta > 0$, we have:
$$\Pr [ \mbox{\rm \# false positives} \ge 1+\delta] 
  \le {\frac {e^\delta} {(1+\delta)^{(1+\delta)}}}.$$ 
\end{lemma}
\begin{proof}
Let $\alpha_i$ be the $i$th item that is checked through $\sf
BF.Check$. That is, we consider a sequence of $${\sf
BF.Check}(\alpha_1), \ldots, {\sf BF.Check}(\alpha_{m}),$$
where $\alpha_i$ is an arbitrary item. 
Since we wish to upper bound the false positives (i.e., we don't care about true
positives), it suffices to consider the
case that for every $i$, 
$\alpha_i \not \in {\sf BF}$ (i.e, $\alpha_i$ has not been inserted in the
BF) as this maximizes the number of possible false positives.

Let $X_1, \ldots, X_m$ be independent Bernoulli random variables with
$\Pr[X_i = 1] = 1/m$. 
Since the BF false positive rate is assumed to be $1/m$, we have
for all $i$, 
$$\Pr[{\sf BF.Check}(\alpha_i)=1] = \Pr[\mbox{query $i$ is a false positive}] \le
1/m.$$
Thus, we can bound the number of false positives
by $\sum_{i=1}^m X_i$.

Now, let $\mu :=
\Exp[\sum X_i] = m \cdot \frac 1 m = 1$.  By applying the Chernoff bound with $\mu =
1$,  we have: 
$$ \Pr\left[\sum_{i=1}^m X_i \ge 1 + \delta \right]  
\le {\frac {e^\delta} {(1+\delta)^{(1+\delta)}}} .$$
\end{proof}

\section{Proof of Theorem~\ref{thm:ssp-code}}\label{app:proof63}

\begin{theorem}[6.3]
Given an FHE scheme, and a $(n, s, c, f_p)$-CODE scheme over domain $D$
in the random oracle model, the construction in Algorithm~\ref{alg:ssCODE} yields a $(\ell,
f_p)$-secure search scheme for records in domain $D$ in the random oracle
model, where $\ell = \frac{c(s)\cdot \ell_c}{s}$, $\ell_c$ is the length
of an FHE ciphertext with plaintext space $D$, and $s$ is the number of
matching records.
\end{theorem}

\begin{proof}
We begin by proving that the adversary cannot distinguish
between two different queries.
The adversary chooses a database $x$ and two queries $q^0, q^1$, with
the promise that $s = \sum_{i=1}^n q^0(x_i) = \sum_{i=1}^n q^1(x_i)$.

The entire view of the adversary during the experiment can be
reconstructed efficiently given (1) the encrypted database $\lbra x
\lket$, (2) the encrypted query $\lbra q \lket$, (3) the decrypted value
of $s$.

We note that the CODE scheme may return either more than $s$ values to the client (in case of a false positive) or less than $s$ values (in case decoding fails), but both of these occur with probability at most $\negl(\secparam)$ and thus we can ignore them in the following.

Since the value of $s$ is the same for $q_0$ and $q_1$, the only thing
that changes in the view of the adversary when switching from $b = 0$ to
$b=1$ is the encrypted query $\tilde{q_b}$.  Therefore, the adversary
guesses $b$ with negligible advantage by the IND-CPA security of the FHE
scheme.

The proof that the adversary cannot distinguish between the same query
applied to two different databases follows nearly identically.

\end{proof}